\input epsf
\documentstyle[proceedings]{crckapb}
\def\msol{M_\odot}
\def\lsol{L_\odot}
\def\eso{ESO\,439-26\,}
\def\te{T_{eff}}
\def\mv{M_V}
\def\mbol{M_{bol}}
\def\wig#1{\mathrel{\hbox{\hbox to 0pt{%
\lower.5ex\hbox{$\sim$}\hss}\raise.4ex\hbox{$#1$}}}}

\begin{opening}

\title{Cool white dwarfs : cooling theory and galactic implications}

\author{G. Chabrier}

\institute{C.R.A.L.,
Ecole Normale Sup\'erieure, 69364 Lyon Cedex 07, France\\
chabrier@ens-lyon.fr}

\end{opening}
\begin{document}

\section{Introduction}

The understanding of the physics of cool white dwarfs (WD) bears important
consequences for Galactic evolution and cosmological implications. The
observed cutoff in the disk WD luminosity function (WDLF)
yields the determination of the age of the Galactic disk, as suggested
initially by Winget et al. (1987). The recent microlensing observations
toward the LMC (Alcock et al., 1996) suggest that WDs might provide a substantial fraction of the halo dark matter (Chabrier, Segretain \&
M\'era, 1996; Adams \& Laughlin, 1996). The correct analysis of these
applications implies a correct WD cooling theory and reliable
photometric predictions, which in turn require accurate interior and
atmosphere models. Important improvement in this latter domain has been accomplished recently by Bergeron, Saumon \&
 Wesemael (1995) and Bergeron, Wesemael \& Beauchamp (1995), which yields
the determination of photometric color indices and bolometric corrections down to 4000 K (see Leggett, these proceedings).
In this paper, we review the most recent improvement in WD interior and cooling theory.

\section{Internal structure. Cooling theory}

The theory of WD cooling was first outlined by Mestel \& Ruderman (1967)
who identified the dichotomic properties of WDs, where degenerate
electrons provide the pressure support but do not contribute significantly
to the heat capacity ($c_V/k_B= \pi^2/2(T/T_F)$ for a fermion gas, where $c_V$
is the specific heat per unit mass and $T_F\sim 10^9$ K is the electron Fermi temperature for WD central densities) while the
ions contribute negligible pressure but provide most of the thermal energy.
Abrikosov (1960) and Salpeter (1961) independently pointed out the possible
onset of crystallization in cool WDs.
VanHorn (1968) first developed a
 consistent theory of WD crystallization and Lamb \& VanHorn (1975) first calculated the evolution of a pure carbon crystallizing WD. These calculations
 have been extended to C/O mixtures by Wood (1992) who examined extensively
 the importance of the various parameters entering WD evolution, e.g. the
 core composition and the atmosphere structure. A further significant breakthrough
 in WD cooling theory is due to Stevenson (1980) who first pointed out
 the importance of the crystallization diagram in a {\it
two-component} (e.g. C/O) plasma and of the difference of chemical composition
 in the fluid and in the solid phase. This motivated numerous calculations for
the characterization of the phase diagram (Barrat, Hansen \&
 Mochkovitch, 1988; Ichimaru, Iyetomi \& Ogata, 1988), the
 effect on WD cooling (Mochkovitch, 1983) and on the WDLF (Garc\' ia-Berro,
 Hernanz, Mochkovitch \& Isern, 1988). More recently Segretain \& Chabrier (1993)
 characterized the evolution of the crystallization diagram of stellar
 plasmas for arbitrary binary mixtures as a function of the charge ratio.
 The chemical differentiation at crystallization calculated with these
 diagrams was shown to produce an extra source of energy $\Delta E$ in the WD (Segretain et al., 1994),
 which in turn leads to a substantial increase in the age of crystallized
 WDs for a given luminosity ($\Delta t \propto \Delta E/L$), an important
 issue for a correct determination of the age of the Galactic disk (Hernanz al., 1994).
 
 In this section, we derive an analytical theory for the evolution of
 cool, crystallizing WDs, based on first principles of thermodynamics
 (Landau \& Lifschitz, 1980), aimed at describing the main physical effects
 in terms of simple physics (see also Isern et al., 1997). These calculations include all the afore-mentioned
 processes and yield a reasonable estimate of the gravitational energy
 release and time delay induced by chemical fractionation at crystallization,
 a question of strong debate among the community. This provides
 useful guidelines to verify the validity of complete,
 numerical calculations.

 The first laws of thermodynamics yield for white dwarf cooling
:

 \begin{eqnarray}
L=-\int_0^M {dq\over dt}dm\,-\,\int_0^M \epsilon_\nu dm=
-\int_0^M {du + d\Omega \over dt}dm\,-\,\int_0^M \epsilon_\nu dm
\end{eqnarray}

where $\epsilon_\nu$ is the neutrino rate and $dq/dt$ is the
 heat rate per unit mass along the change of an equilibrium state. $du$ and $d\Omega$ are the change of specific internal energy and gravitational energy, respectively. The first one reads :

\begin{eqnarray}
du & = & c_VdT \,+\,\bigl( (T{dP\over dT})_{_V}-P \bigr)dv\,+\,\Delta u_{crys}\,+\,\bigl[\Sigma_i(\mu_idN_i)_l\,+\,\Sigma_i(\mu_idN_i)_s\bigr]_{_{V,T}}\nonumber \\
 & = & Tds\,-\, Pdv\,+\, T\Delta s\,+\,
\bigl[\Sigma_i(\mu_idN_i)_l\,+\,\Sigma_i(\mu_idN_i)_s\bigr]_{_{V,T}}
\end{eqnarray}

The $\mu_i$ denote the chemical potentials, $s$ the specific entropy and
$v=1/\rho$ is the volume per unit mass. The $d N_i$ are the variations of carbon and oxygen nuclei in the fluid and the solid phase due to the change of composition at crystallization. They can be calculated with the thermodynamics lever rule. When the central crystal grows\footnote{Note that this fractionation process is described sometimes in the literature as drowning O-{\it flakes}.
This is an erroneous picture, based on Stevenson's old {\it eutectic} diagram, which
corresponds to an {\it inhomogeneous} solid. The
correct {\it spindle} diagram yields a homogeneous solid. It is
easy to verify that the solid is denser than the
liquid, so that the crystal grows at the center of the star. Were WDs made of water, the story would be different because of the volume expansion at crystallization}, there is a thin
C-enriched
surrounding fluid layer where {\it locally} the variation of nuclei
$d(\delta N_{i_l})$ is {\it not} equal to $-d
 N_{i_s}$ (no {\it local} mass conservation) \footnote{This is similar
to e.g. silicium deposition for semi-conductor devices in a silicium+impurity
liquid, which yields a silicium concentration gradient and eventually a silicium-rich layer. We have in fact a distillation process in the WD.}. How this carbon excess in the fluid
is redistributed homogeneously will be examined below. No variation of composition yields of course $\delta N_i=0$ and thus no extra internal energy.

 The first two terms on the r.h.s. of (2) have been recognized originally by Mestel \& Ruderman (1967),
 the third term was first introduced by Van Horn (1968) and is the crystallization
latent heat $l=-\Delta u_{crys}=T(s_{sol}-s_{liq})$.

The condition of hydrostatic equilibrium yields for the variation of gravitational energy :

\begin{eqnarray}
\delta \Omega=\int_0^M P \delta v \delta m\approx \,<{P_e\over \rho}> M
\end{eqnarray}

\noindent since the electron pressure largely dominates the ionic pressure.

 Eqns. (1)-(3) can be rewritten :

 \begin{eqnarray}
L+L_\nu=-\int_0^Mc_V{dT\over dt}dm\,-\,\int_0^M(T{dP\over dT})_{_V}
 {dv\over dt}dm\,+\,l{dm_S\over dt}\,+\delta u{dm_S\over dt}
\end{eqnarray}

 where $\delta u= \Sigma_i (\int \mu_i d(\delta N_i))$ and $m_S$ is the mass crystallized. Note that the
 last two terms in eqns. (2) and (4) are evaluated at {\it constant volume}
and thus {\it do not stem from a contraction work, but from the change of composition at crystallization}.

 The contribution of the first three terms of eqn. (4) can be estimated
 easily, as done initially by Mestel \& Ruderman (1967) and Lamb \& Van Horn (1975) :

\begin{eqnarray}
\delta U_{th} & = &\int_0^Mc_V dT \, dm \approx \, \delta \Omega \,\,\mbox{ from the virial theorem} \nonumber \\
\delta U_{grav} & = & \int_0^MT{dP\over dT}\,dvdm \approx \int_0^M [T{dP_i\over dT}\,+\, o(T/T_F)^2]dvdm
\approx <{P_i^{th}\over \rho}>M \nonumber \\
\end{eqnarray}

where $P_i^{th}=\rho {\cal R}\,T/\mu_0$ (where $\mu_0$ is the mean ionic molecular weigth) is the thermal (non electrostatic) ionic pressure.

Equations (3) and (5) yield $\delta U_{grav} \sim {<P_i^{th}>\over <P_e>}\,\delta \Omega$.
WD characteristic central density $\rho \sim 10^6$ g.cm$^{-3}$ and central temperature $T\sim 10^6$ K yield
$P_i^{th}/P_e\sim 10^{-3}/\bar Z$.
Thus only a negligible fraction of the energy due to gravitational contraction is radiated. Most of the work is expended in raising the electron Fermi energy, as first noted by Lamb \& Van Horn (1975).
The latent heat contribution can be estimated from the differences between the solid and liquid ionic entropies (VanHorn, 1968) :
$l \sim -kT_c/AH$,
where $T_c$ is the crystallization temperature ($\sim 3\times 10^6$ for C/O, see Segretain \& Chabrier, 1993), $A$ is the mean atomic mass
and $H=1$ a.m.u.$=1.66\times 10^{-24}$ g. This yields an energy release
$ U_{latent\, heat} \sim 10^{47}$ erg $\sim 10^{-2}\,\Omega$, where
$\Omega = GM^2/R\sim 10^{49}$ erg is the WD gravitational energy. The negative sign indicates that
the energy is {\it emitted} at crystallization.

The last term can be estimated as follows :

\begin{eqnarray}
\delta u=({\partial u\over \partial X})_{_V,T}\delta X \approx ({\partial u_i\over \partial X})_{_V,T}\Delta X \approx \Delta u_i
\end{eqnarray}

where $X$ is the mass fraction of one of the
 components (say carbon), $\Delta X=X_l-X_s$ and $\Delta u_i= u_{i_l}-u_{i_s}$ is the difference of Madelung energy in
the C/O plasma between the fluid and the solid phase\footnote{In fact $\delta u$ is only a fraction of $\Delta u_i$ since it stems from the difference w.r.t.
to the average energy over the C-enriched layer. This is not
consequential for the present estimate}. The Madelung energy (per unit mass)
of the mixture reads :

\begin{eqnarray}
 u_i/kT = \alpha \Sigma_k{X_k\over A_k} \Gamma_k=\alpha \Gamma_e\bigl\{ X{Z_1^{5/3}\over A_1}+(1-X){Z_2^{5/3}\over A_2}\bigr\}
\end{eqnarray}

where $\alpha$ denotes the Madelung constant in the fluid or in the solid phase
($\alpha_s=-0.9, \alpha_l=-0.899$),
$\Gamma_e=e^2/a_ekT$ ($a_e$ is the mean inter-electronic distance) and the index $k$ denotes each ionic species ($C^{6+},O^{8+}$).
This yields:

\begin{eqnarray}
 \Delta u_i/kT\approx -0.9 \Gamma_e \, \Delta X(Z_1^{5/3}/A_1-Z_2^{5/3}/A_2)
\end{eqnarray}

The virial theorem $P_i/\rho={1\over 3} u_i$ yields :

\begin{eqnarray}
\delta U/\delta \Omega={\int \delta u\, dm_s \over \delta \Omega}
\sim {<\Delta P_i>\over <P_e>}
{M_s\over M}={\Delta P_i\over P_i}{P_i\over P_e}{M_s\over M}
\end{eqnarray}

Note that $P_i$ is now the ionic electrostatic pressure.
For $Z_1=6, Z_2=8$, $x_1$=$x_2$=1/2, we get $P_i/P_e\sim 10^{-2}-10^{-1}$, $\Delta P_i/P_i\sim 2 \Delta x \,(Z_1^{5/3}-Z_2^{5/3})/(Z_1^{5/3}+Z_2^{5/3})\sim
-0.5\times \Delta x$, where $\Delta x\sim 0.1-0.3$ is the difference of carbon number concentration between the solid and the fluid phase at crystallization (Segretain \& Chabrier, 1993). With $M_s/M\sim 0.1$, this yields:

\begin{eqnarray}
\delta U/\delta \Omega \sim 10^{-4}-10^{-3}
\end{eqnarray}

in agreement with the detailed numerical calculations
(see Figure 5 of Segretain et al., 1994).

Chemical differentiation at crystallization thus provides an additional source of energy wich remains much smaller than the gravitational energy. But, as shown
below, the release of this quantity at a low-luminosity phase of the evolution
has a significant effect upon the lifetime of the star at these stages.





\begin{eqnarray}
\Delta t = \int_0^M {\delta u(T)\over L(T)} dm \approx {\delta U(T)\over L(T)} \approx {\Delta P_i\over P_i} {P_i\over P_e} {M_S\over M} {\Omega\over L}
\end{eqnarray}

With the afore-mentioned values we get $\Delta t \sim 5\times 10^{8}$ yr {\it at the begining
of crystallization}, $M_s/M\sim 10\%$ and $L=10^{-3.5}\,L_\odot$ and
$\Delta t \sim 2\times 10^{9}$ yr at $L=10^{-4.5}\,L_\odot$, $M_s/M\sim 80\%$,
the observed cutoff luminosity.
These simple calculations show the importance of the
time delay induced by chemical fractionation at crystallization for cool
(faint) WDs, even though the corresponding energy is small compared to the binding
energy.

An other issue concerns the redistribution of the excess of carbon in the fluid
at crystallization since the solid core is O-enriched, as obtained from the
phase diagram. Since the fluid C-enriched layer around the crystal is lighter than the surrounding medium, a Rayleigh-Taylor instability develops locally, due to the variation of molecular weight.
This problem has been considered in detail by Mochkovitch (1983) who showed that the typical crystallization time (for a 0.6$\msol$ WD, the crystallization velocity at the begining of crystallization is $v_c\sim 10^{-2}\,M_{WD}/7\times 10^7 \,yr\sim 10^{-8}$ cm.s$^{-1}$) is significantly larger than the convection time so that the liquid is likely
to be rehomogeneized rapidly as crystallization goes on.

\section{Galactic implications}

The first application of these calculations concerns the age of the faintest WD ever observed, $ESO\,439-26$ (Ruiz et al., 1995).
The trigonometric-parallax determination of this object yields an absolute magnitude
$\mv=17.4\pm 0.3$, $\mbol=17.1\pm 0.1$ (Bergeron et al., 1997) about 1 mag faintward of the observed cut-off of the WDLF of the Galactic disk (Liebert et al., 1988). The location of this object in the
$\mv$ vs $V-I$ diagram, and a comparison with the photometric sequences of cool white dwarfs
recently derived by Bergeron et al. (1995) yields the interpretation that it is a cool
$(\te \sim 4500$ K), massive WD ($m\sim 1.2 \,\msol$). Comparison with (pure carbon) evolutionary models of Wood (1992)
yields an age determination for this WD, $t\sim 6.5$ Gyr, substantially below the lower limit for the
age of the Galactic disk determined by detailed WD cooling theory (Hernanz et al., 1994).
The Wood sequences do not include the afore-mentioned release of gravitational energy due to C/O
differentiation at crystallization.
Figure 1 displays several isochrones of a C/O WD for several masses
(WDs more massive than $1.2\,\msol$ have a different O/Ne/Mg internal composition), obtained with a WD cooling theory
including (solid line) and neglecting (dashed line) the fractionation process. 
The dotted lines display the luminosity of \eso
 and its mass, assuming either
a H-rich ($\sim 1.1\,\msol$) or a He-rich ($\sim 1.2\,\msol$) atmosphere (see Ruiz et al., 1995). As shown,
\eso
is compatible with an
age t$\sim 10$ Gyr, in good agreement with the most recent determination
of the age of the disk, whereas neglecting differentiation would yield $t< 9$ Gyr.

\begin{figure}
\epsfxsize=100mm
\epsfysize=70mm
\epsfbox{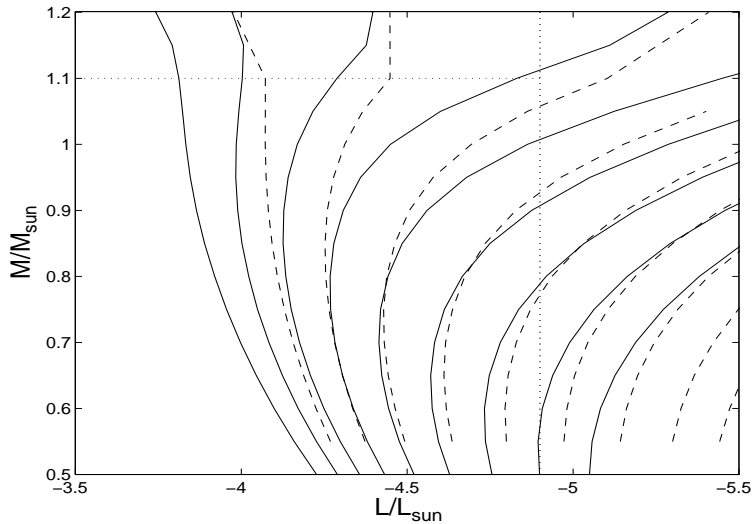}
\caption{Isochrones for different masses of crystallized WDs ($\log L/\lsol < -3.5$) in a mass-luminosity diagram. The solid lines include C/O differentiation at crystallization whereas the dashed lines do not. The isochrones are the {\it same in both cases} and go from $t=7$ Gyr to $t=15$ Gyr from left to right. The vertical dotted line is the inferred luminosity for $\eso$ (Bergeron et al., 1997), while the horizontal dotted lines are the inferred mass for a H-rich
(1.1 $\msol$) and He-rich (1.2 $\msol$) atmosphere. Crystallization starts earlier for massive (and thus rapidly evolving) WDs, which then enter the rapid Debye cooling regime, which causes the bending in the isochrones. }
\end{figure}

An other important galactic application is the age of the disk inferred from the comparison between the observed and the theoretical WDLF. As shown by Hernanz et al. (1994) and by the present analysis, chemical fractionation yields 1 to 2 Gyr
older ages (i.e. 10-20\% increase), depending on the initial C/O profile, w.r.t. estimates which do not include this process.
Preliminary calculations along the present lines based on the recently
observed WDLF (Oswalt et al., 1996) yield an
estimate for the age of the cutoff at $\log L/\lsol=-4.5$, i.e. an
age for the Galactic disk, $t_D\sim 12$ Gyr.

\section{Uncertainties in the theory}

An important uncertainty in the models is the initial C/O composition and stratification. If
the initial WD is already stratified, with oxygen accumulated near the center,
as suggested by Mazzitelli \& D'Antona (1986), the energy release by differentiation, and the related time delay are reduced (see Segretain et al, 1994, Tables 3-4). The other main source of uncertainty concerns the model atmospheres and most importantly for the age
of the disk, the thickness of the helium layer wich regulates the emergent heat flux. The thicker the layer the younger the age (Wood, 1992). These uncertainties translate in a $\sim 1-2$ Gyr uncertainty on the age of the disk.
Futher progress in this direction is certainly essential for
a better determination of the cooling and the observational properties of cool
WDs.

\section{Conclusion}

We have shown in this review that substantial improvement in the theory of
cool WDs has been accomplished within the past few years for the atmosphere
as well as for the internal structure. We have shown that chemical fractionation
at crystallization, although it liberates a negligible amount of gravitational
energy, modifies substantially the cooling history of the star and yields an
important time-delay for faint WDs. This process cannot be ignored in accurate
WD cooling theory.
As noted by Mochkovitch (private communication), although the importance of crystallization of
alloys in stellar plasmas (WDs in particular) is still strongly debated in the
astrophysical community, it has been recognized long ago in geophysics
(see e.g. Loper, 1984; Buffett et al., 1992). Although the nature of the plasma
is different, the {\it physics} of the process (thermodynamics and energy transport) is exactly the same.
Important uncertainties remain essentially in the exact determination of the initial C/O profile in the star and in the structure of
the hydrogen and/or helium outer envelope.


\small
\tt\raggedright
\parindent=0pt
\medskip
{\bf DISCUSSION}

\medskip

ROBERT KURUCZ: What about the effect of rotation on solidification?

\medskip

GILLES CHABRIER: It is not clear how rotation will affect the crystal growth, but there is no reason for it to {\it prevent} crystallization, which must occur below a certain temperature. The effect of rotation on the C-redistribution in the surrounding layer has been considered in detail by Mochkovitch (1983).


\begin{thebibliography}{}

\bibitem[{Abrikosov}]{A61}
Abrikosov, A.A., 1961, J.E.T.P., 12, 1254

\bibitem[{Adams}]{A61} 
Adams, F. \& Laughlin, G., 1996, ApJ, 468, 586

\bibitem[{Alcock96}]{A61}
Alcock, C. et al., 1996, astroph-9604176


\bibitem[{BHM}]{A61}
Barrat, J.L., Hansen, J.P. \& Mochkovitch, R., 1988, A\&A, 199, L15



\bibitem[{Bergeron {\it et al.}}{1995a}]{Ber95a}
Bergeron, P., Saumon, D., \& Wesemael, F, 1995, ApJ 443, 764

\bibitem[{Bergeron {\it et al.}}{1995b}]{Ber95b}
Bergeron, P., Wesemael, F. \& Beauchamp, A., 1995, PASP, 107, 1047

\bibitem[{Buffett}]{A61}
Buffett, B.A., Huppert, H.E., Lister, J.R. \& Woods, A.W., 1992, Nature, 356, 329

\bibitem[{Chabrier et al.}{1996}]{Cha96}
Chabrier, G., Segretain, L. \& M\'era, D., 1996, ApJ, 468, L21

\bibitem[{Dubin}]{A61}
Dubin, D., 1990, Phys. Rev. A, 42, 4972

\bibitem[{Garcia}]{A61}
Garc\' ia-Berro, E., Hernanz, M., Isern, J. \& Mochkovitch, R., 1988, A\&A, 193, 141

\bibitem[{Hernanz}]{A61}
Hernanz, M., Garc\' ia-Berro, E., Isern, J., Mochkovitch, R., Segretain, L. \& Chabrier, G., 1994, ApJ, 434, 652

\bibitem[{Ichi}]{A61}
Ichimaru, S., Iyetomi, H. \& Ogata, S., 1988, ApJ, 334, L17

\bibitem[{Isern}]{A97}
Isern, J., Mochkovitch, R., Hernanz, M., Garc\' ia-Berro, E., 1997, ApJ, submitted



\bibitem[{LVH}]{A61}
Lamb, D.Q. \& VanHorn, H.M., 1975, ApJ, 200, 306

\bibitem[{LL}]{A61}
Landau, L. \& Lifshitz, E., 1980, {\it Statistical Physics}, Pergamon Press

\bibitem[{Loper}]{Lo84}
Loper, D.E., 1984, Adances in Geophysics, 26, 1

\bibitem[{LDM}]{A61}
Liebert, J., Dahn, C.C.\& Monet, D.G., 1988, ApJ, 332, 891

\bibitem[{MD}]{A61}
Mazzitelli, I. \& D'Antona, F., 1986, 308, 706 

\bibitem[{MR}]{A61}
Mestel, L. \& Ruderman, M.A., 1967, M.N.R.A.S., 136, 27

\bibitem[{Mochko}]{A61}
Mochkovitch, R., 1983, A\&A, 122, 212

\bibitem[{Oswalt}]{A61}
Oswalt, T.D., Smith, J.A., Wood, M.A. \& Hintzen, P., 1996, Nature, 382, 692

\bibitem[{Ruiz}]{A61}
Ruiz, M.T., Bergeron, P., Leggett, S., \& Anguita, C, 1995, ApJ, 455, L159

\bibitem[{Sal}]{A61}
Salpeter, E.E., 1961, ApJ, 134, 669


\bibitem[{SC}]{A61}
Segretain, L. \& Chabrier, G., 1993, A\&A, 271, L13

\bibitem[{Seg}]{A61}
Segretain, L., Chabrier, G., Hernanz, M., Garc\' ia-Berro, E., Isern, J. \& Mochkovitch, R., 1994, ApJ, 429, 641


\bibitem[{Stevenson}]{A61}
Stevenson, D., 1980, J. Physique Sup., 41, C2-61

\bibitem[{VH}]{A61}
VanHorn, H.M., 1968, ApJ, 151, 227

\bibitem[{Winget}]{A61}
Winget, D. et al., 1987, ApJ, 315, L77

\bibitem[{Wood}]{A61}
Wood, M. A., 1992, ApJ, 386, 536


\end{thebibliography}
\end{document}